\documentclass[aps, prd, preprint, amsfonts, floatfix]{revtex4}
\usepackage{graphicx}
\usepackage{amsmath,amsfonts}
\usepackage{epsfig,color}

\begin{document}
\newcommand{\be}{\begin{eqnarray}}
\newcommand{\ee}{\end{eqnarray}}
\newcommand\del{\partial}
\newcommand\nn{\nonumber}
\newcommand{\Tr}{{\rm Tr}}
\newcommand{\Str}{{\rm Trg}}
\newcommand{\mat}{\left ( \begin{array}{cc}}
\newcommand{\emat}{\end{array} \right )}
\newcommand{\vect}{\left ( \begin{array}{c}}
\newcommand{\evect}{\end{array} \right )}
\newcommand{\tr}{{\rm Tr}}
\newcommand{\hm}{\hat m}
\newcommand{\ha}{\hat a}
\newcommand{\hz}{\hat z}
\newcommand{\hze}{\hat \zeta}
\newcommand{\hx}{\hat x}
\newcommand{\hy}{\hat y}
\newcommand{\tm}{\tilde{m}}
\newcommand{\ta}{\tilde{a}}
\newcommand{\U}{\rm U}
\newcommand{\diag}{{\rm diag}}
\newcommand{\tz}{\tilde{z}}
\newcommand{\tx}{\tilde{x}}
\definecolor{red}{rgb}{1.00, 0.00, 0.00}
\newcommand{\rd}{\color{red}}
\definecolor{blue}{rgb}{0.00, 0.00, 1.00}
\definecolor{green}{rgb}{0.10, 1.00, .10}
\newcommand{\blu}{\color{blue}}
\newcommand{\green}{\color{green}}
\newcommand{\omegat}{\tilde \omega}


\title{Distribution  of Canonical Determinants in  QCD}

\author{Andrei Alexandru}
\affiliation{Physics Department, The George Washington University, Washington DC, USA}

\author{C. Gattringer, H.-P.~Schadler}
\affiliation{Institut f\"ur Physik, FB Theoretische Physik, Universit\"at Graz, 8010 Graz, Austria}

\author{K. Splittorff}
\affiliation{Discovery Center, The Niels Bohr Institute, University of Copenhagen, 
Blegdamsvej 17, DK-2100, Copenhagen {\O}, Denmark} 

\author{J.J.M. Verbaarschot}
\affiliation{Department of Physics and Astronomy, Stony Brook University, Stony Brook,
 New York 11794, USA}

\date   {\today}
\begin  {abstract}
The distribution  of canonical determinants in QCD 
is determined by means of  chiral perturbation theory. For a non-zero quark charge the
canonical determinants take complex values. 
In the dilute pion gas approximation, we compute all moments of
the magnitude of the canonical determinants, as well as the first
nonvanishing moments of the real and imaginary parts. The non-trivial
cancellation between the real and the imaginary parts of the canonical
determinants is derived and  the
signal to noise ratio is discussed. The analytical distributions are
compared to lattice data. The average density  of the magnitude
of the canonical determinants is determined as well and is shown to
be given by a variant of the log-normal distribution.

\end{abstract}
\maketitle

\section{Introduction}
QCD at finite baryon density and low temperatures is one of the 
least understood regions of the QCD phase diagram. The reason is that 
the baryon chemical potential introduces a sign problem which invalidates standard stochastic methods to
evaluate the path integral for the QCD 
partition function. 
The source of the problem 
goes back to the fluctuations of the pions
when the chemical potential exceeds half the pion mass. Then
the value of the average 
fermion determinant is strongly suppressed with respect to 
the average of the magnitude of the fermion determinant (for reviews 
and additional details see
\cite{deForcrand:2010ys,Splittorff:2006vj,Splittorff:2007zh,Splittorff:2007ck,Splittorff:2006fu}).
One way to evade this problem might
be to use the canonical ensemble \cite{haas,deforcrand,liu,DG}. 
This approach only requires the QCD partition function at 
imaginary chemical potential which can be calculated reliably by standard
methods \cite{roberge,alford,mario,philip}.
However, the extraction of  the canonical partition function requires 
the evaluation of  the Fourier transform
\be
Z^q = \frac 1 {2\pi} \int_{-\pi}^\pi d\theta\; e^{-iq\theta} Z(\mu= i\theta T),
\ee
which in the thermodynamic limit, $q \to \infty$, leads to unmanageable
cancellations. In lattice simulations, one can do better though. The
fermion determinant at non-zero chemical potential is a polynomial
in $\exp[\pm \mu/T] $,
\be
Z(\mu) = \langle \sum_{q=-N}^N D_q \; e^{\mu q/T} \rangle,
 \ee
so that $Z^q =\langle D_q \rangle$. Because this is a finite polynomial the Fourier
coefficients can be determined exactly by means of a discrete Fourier
transform.  The  computation of  the average of the canonical determinants, 
however, remains 
a challenge.
In this paper we show that this challenge depends crucially on the value of $q$. For small
$q$ the signal to noise ratio is tractable while it becomes exponentially small 
with increasing $q$.

To address the signal to noise problem in the canonical approach
 we will evaluate the average magnitude of the canonical determinants 
to one-loop order in chiral perturbation theory and compare them to the 
average of the canonical determinants. This will give us information
on the degree of cancellations that take place in evaluating the canonical
partition function. The magnitude is obtained from the absolute value
of the canonical determinants. This carries an isospin 
component which couples to the pions. On the contrary 
the charge dependence of the average of the canonical determinants themselves 
is not directly coupled to the pions. This difference leads to an  
exponentially small (in the charge)  signal to noise ratio.  
We also obtain the full distribution of the absolute value
of the canonical determinants and make comparisons with lattice results for 
the $q$-dependence  of the canonical determinants.

The problems facing the canonical approach have also been emphasized in 
\cite{kaplan}. Though the entire approach is different, the problems were 
traced back to the same source, namely that the average value of
the baryon correlator is strongly suppressed with respect to the ``noise'' 
given by the absolute value squared correlator which at large distances is
dominated by the contribution of the pions.

We start this paper with a discussion of canonical partition functions
 corresponding to non-zero isospin chemical potential (section 2). 
In section 3 we show that  
the same expressions give the magnitude of the canonical determinants
at non-zero baryon number density. The relative size of the cancellations
which take place in the evaluation of the canonical partition function
is estimated in section 4 by modeling the nucleon contribution in terms of
 a resonance gas model. In section 5 we compare the $q$-dependence
of the canonical partition functions with results from lattice QCD simulations.
The full distribution of the magnitude of the canonical determinants is derived in section 6 and concluding remarks are made in section 7. Additional details 
are worked out in three appendices.

\section{Canonical partition functions for isospin charge}

Before we turn to the distribution of the canonical determinants in QCD with
non-zero quark charge it is useful to compute the canonical 
partition functions at fixed isospin number.
\vspace{2mm}

In oder to derive the canonical partition function with isospin charge we
first recall the relation between the grand canonical partition function and
the canonical partition function. For simplicity we consider two light flavors
of mass $m$.

The two flavor QCD partition function at non-zero isospin chemical
potential $\mu$ is given by 
\be
\label{Z11*}
&& Z_{1+1^*}(\mu) = \langle{\det}(D+\mu\gamma_0+m) \det(D-\mu\gamma_0+m)\rangle,
\ee
where $D$ is the Dirac operator and  the average is over the Yang-Mills action. 
This grand canonical partition function
can be decomposed in terms of canonical partition functions as
\be
Z_{1+1^*}(\mu)=\sum_{q=-\infty}^\infty e^{q\mu/T}Z_{1+1^*}^q,
\ee
with 
\be
Z_{1+1^*}^q = \frac{1}{2\pi}\int_{-\pi}^\pi d\theta \ e^{-i\theta q} 
Z_{1+1^*}(\mu/T=i\theta).
\label{Ftrans-muI}
\ee

We evaluate the canonical partition functions to one-loop order in
chiral perturbation theory. At this order, the grand canonical partition function in
the normal phase is given by \cite{Splittorff:2007ck}
\be
\label{Z11*-normal}
&& Z_{1+1^*}(\mu) = \langle{\det}(D+\mu\gamma_0+m)
        \det(D-\mu\gamma_0+m)\rangle \simeq e^{G_0|_{V=\infty}+g_0(\mu)}, 
\ee
where the $\mu$ dependence of the free energy resides entirely in the
part (see \cite{STV})
\be
g_0(\mu) =\frac{Vm_\pi^2T^2}{\pi^2}\sum_{n=1}^\infty\frac{K_2(\frac{m_\pi n}{T})}{n^2}\cosh(\frac{2\mu n}{T}).
\ee
The $\mu$ independent part, $G_0|_{V=\infty}$, will be suppressed throughout 
(the final results will be ratios where this contribution drops out).
The corresponding 
canonical partition functions are given by the coefficients of $\exp(\mu q/T)$ in the
expansion
of the one-loop result for the $\mu$-dependent part of 
the  grand canonical partition function
in powers of $\exp(\mu/T)$
\be
Z_{1+1^*}(\mu)&=&  
 Z_{1+1^*}(\mu=0)\; \exp\left({\frac{Vm_\pi^2T^2}{\pi^2}\sum_{n=1}^\infty\frac{K_2(\frac{m_\pi n}{T})}{n^2}
(\cosh(2\mu n/T)-1)}\right)\nn\\
&=& \sum_q e^{q\mu/T}Z^q_{1+1^*}. 
\label{zex}
\ee
Using the asymptotic form of the Bessel function $K_2$ (which corresponds to the non-relativistic limit) we can
make an estimate for the parameter domain where the terms with $n > 1$ can be ignored.
From the condition that the correction to the free energy density due to the $n=2$ term
should be much less than the free energy density from the $n=1$ contribution we obtain
\be
\frac {q (2mT)^{3/2}}{V_3}\ll \log\frac {V_3}{q (2mT)^{3/2}}.
\ee
Therefore for a dilute pion gas, 
\be
\frac q{V_3} \ll (2mT)^{3/2},
\label{dilute}
\ee
we  can restrict ourselves to the $n=1$ term. This results in
the canonical partition function
\be
\frac{Z_{1+1^*}^q}{Z_{1+1^*}^{q=0}}  & = & \frac{ e^{-\omega}\int_{-\pi}^\pi \frac{d\theta}{2\pi} \ e^{-i\theta q} e^{\omega\cos(2\theta)}}{ e^{-\omega}\int_{-\pi}^\pi \frac{d\theta}{2\pi} \ e^{\omega\cos(2\theta)}}
\ee
with $\omega$ defined by
\be
\omega = \frac {V_3 m_\pi^2T }{\pi^2}K_2(m_\pi/T).
\ee
For odd values of $q$ the canonical partition functions vanish. This is 
natural since the pions (in the convention used here) carry two units of 
isospin charge. For even $q$ the canonical integral is given by the modified Bessel function
resulting in the ratio
\be
\frac{Z_{1+1^*}^{q} }{Z_{1+1^*}^{q=0} }& = & \frac{I_{q/2}(\omega)}{I_0(\omega)}, \qquad q \ {\rm even}.
\label{I-ratio}
\ee
Thus we have found that 
the distribution 
of the canonical partition functions over $q$
 is described by the modified Bessel functions \footnote{This so called Skellam distribution \cite{skellam} is the distribution of two independent stochastic variables distributed according to Poisson distributions. In our case the Poisson distribution for the pions and anti-pions is given by $\omega^k\exp(\pm 2\mu k/T)/k$ as follows by expanding the exponent in (\ref{Z11*-normal}) for $n=1$.}. 
For $m_\pi/T\gg1$ we can also replace $K_2$ 
by its asymptotic form resulting in
\be
\frac{Z_{1+1^*}^{q} }{Z_{1+1^*}^{q=0} } & = &\frac{I_{q/2}\big(\frac{V_3 m_\pi^{3/2}T^{3/2}}
{\sqrt{2\pi^3}}e^{-m_\pi/T}\big) }
{I_{0}\big(\frac{V_3 m_\pi^{3/2}T^{3/2}}{\sqrt{2\pi^3}}e^{-m_\pi/T}\big)} 
, \qquad q \ {\rm even}.
\ee
The chemical potential corresponding to the canonical partition
function is worked out in \ref{app:mufromZq}.

Finally, it is instructive to sum  over $q$ to obtain the grand canonical
partition function
\be
Z_{1+1^*}(\mu) & = & \sum_{q=-\infty}^\infty e^{q\mu/T}Z_{1+1^*}^q  \\
  & = & Z_{1+1^*}(\mu=0) \sum_{j=-\infty}^\infty e^{2j\mu/T-\omega} I_{j}\Big(\frac{Vm_\pi^{3/2}T^{5/2}}{\sqrt{2}\pi^{3/2}}e^{-m_\pi/T}\Big) \nn \\
   & = & Z_{1+1^*}(\mu=0) \; \exp\left(\frac{Vm_\pi^{3/2}T^{5/2}}{2\sqrt{2}\pi^{3/2}}(e^{(2\mu-m_\pi)/T}+e^{(-2\mu-m_\pi)/T}-2 e^{-m_\pi/T} ) \right), \nn
\ee
which, as it should, brings us back to our starting point.

As we shall see below, the computation of the distribution of the canonical 
{\sl determinants} at non-zero quark charge has many analogies to the above 
computation of the canonical {\sl partition functions} as a function of 
isospin charge. Let us therefore briefly discuss the overall structure of these
computations: 
For the Fourier transform (\ref{Ftrans-muI}) we need the partition function 
at non-zero imaginary isospin chemical potential. For real isospin 
chemical potential less than $m_\pi/2$ the partition function is in the normal phase 
and the expression (\ref{Z11*-normal}) is the 1-loop result from chiral 
perturbation theory for this phase. This leading order expression for 
small real isospin chemical potential is also the leading contribution 
at imaginary isospin chemical potential. For a real isospin 
chemical potential larger than $m_\pi/2$ the partition function is in 
a pion condensed phase. At imaginary isospin chemical potential 
this implies that there is a sub-leading saddle point, which is not taken 
into account above. 
The analytic form of this contribution is worked out using mean field 
chiral perturbation theory  in \ref{condensed}.

\section{The canonical determinants at non-zero quark charge}

Let us now turn to quark chemical potential. Since the pions have zero 
quark charge we obviously have 
\be
Z^{q\neq0}=0
\ee
when evaluated in CPT. However, the fermion determinant 
at non-zero chemical potential can be decomposed
into canonical determinants before averaging over the gauge fields
\be
\det(D+m+\mu\gamma_0) = \sum_q e^{\mu q/T}  D_q,
\ee
with 
\be
D_q = \frac 1{2\pi } \int_{-\pi}^\pi d\theta e^{-i\theta q} 
\det(D+m+i\theta T\gamma_0).
\ee 
Note that  $D_{q=0}$ is real.

Although pions do not have baryon charge, they contribute to the magnitude of $D_q$,
\be 
\langle |D_q|^2\rangle 
&=& \int \frac{d\theta_1 \ d\theta_2}{(2\pi)^2} \ e^{-i(\theta_1-\theta_2)q}
\; \langle \det(D+m+i\theta_1T\gamma_0)\det(-D+m-i\theta_2T\gamma_0)\rangle\nn\\
&=& \int \frac{d\theta_1 \ d\theta_2}{(2\pi)^2} \ e^{-i(\theta_1-\theta_2)q}
\; \langle \det(D+m+i\theta_1T\gamma_0)\det(D+m+i\theta_2T\gamma_0)\rangle.
\label{absdq2}
 \ee
The reason is that 
\be
\langle \det(D+m+i\theta_1T\gamma_0)\det(D+m+i\theta_2T\gamma_0)\rangle
\ee
is the partition function at non-zero imaginary quark, $i(\theta_1+\theta_2)T$, {\sl and} isospin, $i(\theta_1-\theta_2)T$, 
chemical potential. The double Fourier transform in (\ref{absdq2}) singles 
out the contribution with isospin charge $q$ and zero baryon charge. 
To one-loop order in chiral perturbation theory
it is given by
\be
&&\frac{\langle \det(D+m+i\theta_1T\gamma_0)\det(D+m+i\theta_2T\gamma_0)\rangle}
{\langle \det(D+m)^2\rangle}\nn \\
& = &
\frac{ \exp\Big(\frac{Vm_\pi^2T^2}{\pi^2}\sum_{n=1}^\infty\frac{K_2(\frac{m_\pi n}{T})}{n^2}
\cos((\theta_1-\theta_2) n)\Big)}{ \exp\Big(\frac{Vm_\pi^2T^2}{\pi^2}\sum_{n=1}^\infty\frac{K_2(\frac{m_\pi n}{T})}{n^2}\Big)}.
\label{one-loop}
\ee
For $m_\pi/T\gg 1$ when the inequality (\ref{dilute}) is satisfied, 
the average
$\langle |D_q|^2\rangle$ normalized to the $q=0$ expression    simplifies to
\be
\frac{\langle |D_q|^2\rangle}{\langle |D_{q=0}|^2\rangle}
& = &  \frac{\int d\theta_- \ e^{-iq\theta_-}
  e^{\frac{Vm_\pi^2T^2}{\pi^2}K_2(\frac{m_\pi}{T})\cos(\theta_-)} }
{\int d\theta_- \ 
  e^{\frac{Vm_\pi^2T^2}{\pi^2}K_2(\frac{m_\pi}{T})\cos(\theta_-)} }
\nn \\
& = &\frac{I_{q}(\omega)}{I_{q=0}(\omega)}. 
\label{Iq-dist}
\ee
This main result is compared to lattice data in section \ref{sec:lattice}.

In order to address the cancellations in the average of the canonical 
determinant we now evaluate also $\langle D_q^2 \rangle$. This computation 
follows the same lines as above and instead of (\ref{Iq-dist})
we obtain 
\be 
\langle D_q^2\rangle
&=& \int \frac{d\theta_1 \ d\theta_2}{(2\pi)^2} \ e^{-i(\theta_1+\theta_2)q}
\; \langle \det(D+m+i\theta_1T\gamma_0)\det(D+m+i\theta_2T\gamma_0)\rangle.
\label{dq2}
 \ee
Using the one-loop result we obtain for
 $m_\pi/T\gg1$,
\be
\frac{\langle D_q^2\rangle}{ \langle D_{q=0}^2\rangle }
& = &  \frac{\int d\theta_+ d\theta_- \ e^{-iq\theta_+}
  e^{\frac{Vm_\pi^2T^2}{\pi^2}K_2(\frac{m_\pi}{T})\cos(\theta_-)}}{\int d\theta_+ d\theta_- \ 
  e^{\frac{Vm_\pi^2T^2}{\pi^2}K_2(\frac{m_\pi}{T})\cos(\theta_-)}} \nn \\
& = & \delta_{q0}, 
\label{Iq-distb}
\ee
as expected, since the double Fourier transform in (\ref{dq2}) singles out 
the contribution with quark charge $q$ and pions have zero baryon number.

In order to understand better how $\langle D_q^2\rangle=0$ for $q\neq0$ is formed, first note that we have $\langle {D_q^*}^2\rangle = \langle D_q^2\rangle$, which can be rewritten as
\be 
\langle {\rm Re}[D_q] {\rm Im}[D_q] \rangle =0.
\ee
Now, let us express the expectation value of $D_q^2$ in terms of the expectation
value of its real and imaginary parts
\be
\langle D_q^2 \rangle = \frac 14\langle (D_q+D_q^* )^2\rangle + \frac 14\langle (D_q-D_q^* )^2\rangle.
\ee
Next note that both the square of the real part and the square of the imaginary part contain $|D_q|^2$:
\be
\left\langle(D_q^*\pm D_q)^2\right\rangle =  2\left\langle D_q^2 \pm |D_q|^2\right\rangle.
\ee
Therefore, the canonical determinants with non-zero $q$ have  equal variance 
in the real and the
imaginary direction of the complex plane (when evaluated within chiral perturbation theory). These
contributions cancel in the evaluation of $\left\langle D_q^2 \right\rangle$. 

We can also calculate the average value of $|D_q|^2$ from the mean field expression
for the free energy in the condensed phase. The calculation proceeds
along the steps of \ref{condensed} and one obtains the result
\be
\frac{\langle| D_q|^2 \rangle}{\langle| D_{q=0}|^2 \rangle} = 
 e^{-q^2/8 V_4F^2 T^2  -|q|m_\pi/T}  \; .
\ee

\section{Signal to noise ratio}

In the previous section we have seen that 
to one-loop order in chiral perturbation theory 
$\langle D_q^2 \rangle = 0$ for $q \ne 0$. The reason is that in this limit the 
partition function does not contain any baryons. The effect of nucleons can be
taken into account schematically by means of the 
 Hadron Resonance Gas model for $T\ll m_N$ resulting in the ratio
of the two-flavor partition
functions
\be
\frac{Z_q(m_N)}{Z_{q=0}(m_N)} \equiv \frac {\langle D_q^2\rangle}{\langle D_{q=0}^2\rangle}
= \frac{I_{q/3}(\omega_N)}{I_{q=0}(\omega_N)},
\ee
where
\be
\omega_N= \frac {Vm_N^2T^2}{\pi^2}K_2(m_N/T),
\ee
with $m_N$ the nucleon mass \footnote{A similar form of the $q$ dependence of the canonical partition function was found in \cite{BraunMunzinger,Morita}. In \cite{Shinsuke} a Gaussian form was obtained.}.
Therefore the 'signal to noise' ratio of the average  canonical determinants is 
given by (note that $D_{q=0}$ is real and hence that $D_{q=0}^2=|D_{q=0}|^2$)
\be
\frac{\langle D_q^2\rangle}{\langle |D_q|^2\rangle} 
= \frac{I_{q/3}(\omega_N)}{I_{q/2}(\omega)}.
\ee
For large $q$ the ratio $I_{q/3}(Vx)/I_{q/2}(Vy)$ goes as $(x^{1/3}/y^{1/2})^q$.  Using this we find in the dilute limit
\be
\frac{\langle D_q^2\rangle}{\langle |D_q|^2\rangle} 
\simeq e^{-q(m_N/3-m_\pi/2)/T} .
\ee
Note that the 
factors of $V$ cancels. Since $m_N/3>m_\pi/2$ we conclude that at fixed 
baryon density the signal to noise ratio becomes exponentially small in 
the thermodynamic limit.

\section{Lattice Results for Canonical Determinants}
\label{sec:lattice}

The distribution of the canonical determinants has been measured in lattice QCD 
in \cite{DG,BDGLL,AW}. 
In this section we make a first qualitative comparison of the analytical 
prediction for the $q$ dependence of the average magnitude of the canonical 
determinants, Eq.~(\ref{Iq-dist}), to lattice data.

The results we present are for ensembles of $N_f = 2$ Wilson fermions on $12^3\times 6$
lattices generated with the MILC code \cite{MILC}. A hopping parameter of $\kappa = 0.162$
is used and we show results for $\beta = 5.025$ and $\beta = 5.325$, which corresponds to
temperatures of $T = 100 MeV$ and $T = 140$ MeV. The lattice spacing was determined from
the Wilson flow and the pion masses are $m_\pi \sim 900$ MeV. The errors we show are
statistical errors determined with the jackknife method.

\begin{center}
\begin{figure}[t!]
\includegraphics[width=16cm,angle=0]{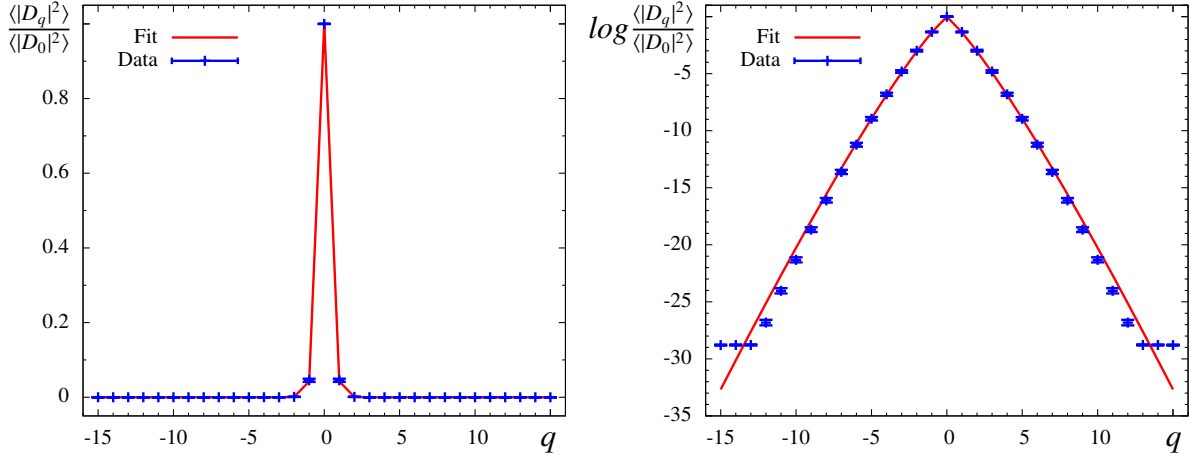}
\caption{\label{fig:lattice1} {\bf Left:} Lattice results for $\langle|D_q|^2\rangle/\langle |D_0|^2\rangle$  (points) at $T=100$ MeV together
 with a fit of $I_q(\omega)/I_0(\omega)$ (solid red curve)
as a function of $q$  (left). {\bf Right:} Same data after taking the logarithm, i.e.~$\log\left(\langle|D_q|^2\rangle/\langle |D_0|^2\rangle\right)$ as a function of $q$. The fit includes the range $-6\leq q\leq6$ and the fitted value is $\omega = 0.085$ (reduced $\chi^2 \simeq 0.4$). The $q$ independence of the data points for $|q|\geq13$ is due to insufficient numerical precision.}
\end{figure}
\end{center}

\begin{center}
\begin{figure}[t!]
\includegraphics[width=16cm,angle=0]{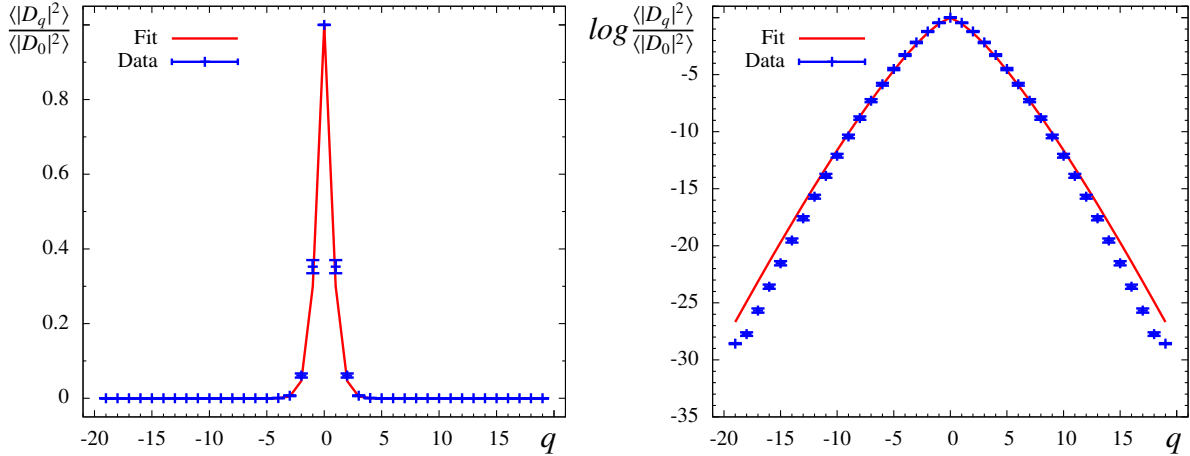}
\caption{\label{fig:lattice2} As in figure \ref{fig:lattice1} but now for $T=140$ MeV: Lattice results for  $\langle|D_q|^2\rangle/\langle |D_0|^2\rangle$ 
({\bf left}) and for  $\log(\langle|D_q|^2\rangle/\langle |D_0|^2\rangle)$ ({\bf right}). 
Again the fit is based on the lower range of $q$, here $-9\leq q\leq9$ ($\omega=0.628$ reduced $\chi^2 \simeq 3.6$), 
where the analytical expression is expected to hold best.}
\end{figure}
\end{center}
\vspace{-1cm}
We have computed 
$\langle |D_q|^2\rangle/\langle |D_0|^2\rangle$ on these lattices and have 
used the argument, $\omega$, of the Bessel $I_q$ functions in Eq.~(\ref{Iq-dist}) 
as a fitting parameter.
In Fig.~\ref{fig:lattice1} we show the results for a temperature of $T=100$ MeV.
The data for $|q|\geq13$ are not to be considered (the accuracy of the Fourier transform 
used does not allow to compute canonical determinants for   $|q|\geq13$). We observe that the fit works well over 20 orders of magnitude. 
We have used 
the data in the range $|q|\leq 6$ for the fit which returns the value 
$\omega = 0.085$, with a reduced $\chi^2 \simeq 0.4$.

Despite the higher temperature the fit for $T=140$ MeV shown in Fig.~\ref{fig:lattice2} is almost as good.
The fitted values of $\omega$ in this case is $\omega=0.628$ with a reduced $\chi^2$  of $3.6$.

\begin{center}
\begin{figure}[t!]
\includegraphics[width=10cm,angle=0]{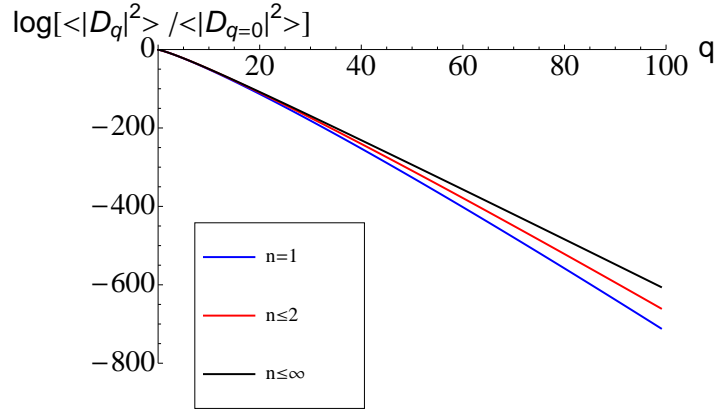}
\caption{\label{fig:n2} The curves  show $\log[\langle |D_q|^2 \rangle/
\langle |D_{q=0}|^2 \rangle$  
 for $V_3 m_\pi^3= 2000$ and $T/m_\pi= 1/6$  (i.e. $ \omega =0.56/\pi^2$),
for $n=1$ (blue), $n\le 2$ (red) and with all values of $n$ included (black).
}
\end{figure}
\end{center}

For larger values of $q$ we find significant deviations between the lattice results
and the analytical fits. 
This could be due to the
 larger $n$ contributions in the ratio of the canonical determinants 
\be
\frac{\langle |D_q|^2\rangle}{\langle |D_{q=0}|^2\rangle} 
& = &  \frac{\int \frac{d\theta_-}{2\pi}
 \ e^{-iq\theta_-}
 e^{\frac{Vm_\pi^2T^2}{\pi^2}\sum_{n=1}^\infty \frac{K_2(\frac{m_\pi n}{T})}{n^2}\cos(n\theta_-)}}{\int \frac{d\theta_-}{2\pi}
 \ e^{\frac{Vm_\pi^2T^2}{\pi^2}\sum_{n=1}^\infty \frac{K_2(\frac{m_\pi n}{T})}{n^2}\cos(n\theta_-)}}.
\ee
Numerically it is no problem to keep more terms in this sum. 
In Fig.~\ref{fig:n2} we compare the $I_q/I_0$ 
distribution (obtained keeping only the $n=1$ term) to the distribution we get keeping 
the $n\le 2$ terms and all terms. As expected the higher $n$ terms 
affect the larger $q$ values more. The effect is of the same magnitude as the deviation 
of the lattice data from the analytical prediction but the sign is opposite.

Because of the rather large pion mass in the present simulation, further 
lattice studies are required in order to quantify the test of the analytic 
expression.

\section{Full Distribution of the Magnitude of the Canonical Determinants}

In this section we will evaluate the average of all moments of $|D_q|^2$ 
in the limit of a dilute 
pion gas and obtain an analytical expression for the density $\rho_q(x)$.
The probability density of the magnitude of the  determinants is given by
\be
\rho_q(x)& = &\langle \delta(|D_q|^2 -x) \rangle \nn\\
&=& \Big\langle \frac 1{2\pi}\int_{-\infty}^\infty ds e^{-is(|D_q|^2 -x)}\Big\rangle \nn\\
&=& \frac 1{2\pi}\int_{-\infty}^\infty ds \sum_{p=0}^\infty\frac{\langle (-is|D_q|^2)^p \rangle}{p!}  
e^{ isx}.
\label{dens}
\ee

Let us first consider the $p$-th moment of the absolute value
squared of the fermion determinant to
one loop order in chiral perturbation theory. The partition function is 
the same as for the theory for $2p$ flavors
and,  to one loop order, each flavor contributes \cite{GL}
\be
-\frac 12 g_0(m_\pi^2,T,L) 
\ee
to the free energy. 
For the $2p$ moment we thus find for vanishing isospin chemical potential
\be
\langle {\det}^{2p}(D+m)\rangle = C e^{\frac 12 (4p^2 -1) g_0(m_\pi,T)},
\ee
which are the moments of a log-normal distribution.

For convenience we use in this section 
a slightly different  normalization of $D_q$ which includes the
contribution of the neutral pions and denote the canonical
determinants by $\tilde D_q$.
 In this normalization
we obtain the moments
\be
\langle |\tilde D_q|^{2p}\rangle
&=& e^{(p-\frac 12)\omega}\int \prod_{k=1}^p \frac{d\theta_k \ d\phi_k}{(2\pi)^2} \\
&&\times  
\prod_{k=1}^pe^{-i(\theta_k-\phi_k)q}
 e^{\frac{Vm_\pi^2T^2}{\pi^2}K_2(\frac{m_\pi}{T})[\sum_{k<l}
\cos(\theta_k-\theta_l)+\cos(\phi_k-\phi_l) +\sum_{k,l}\cos(\theta_k-\phi_l)]}.
\nn
\ee 
where the first exponent is the contribution due to the $2p-1$ neutral pions. 
The second factor in the exponent can be rewritten as
\be
&&\sum_{k<l}[  \cos(\theta_k-\theta_l) + \cos(\phi_k-\phi_l)]
 +\sum_{k,l} \cos (\theta_k-\phi_l) \nn\\ 
&& =
 \frac 12( \sum_{k=1}^p\cos \theta_k  +\sum_{k=1}^p \cos\phi_l)^2 
+\frac 12( \sum_{k=1}^p\sin \theta_k  +\sum_{k=1}^p \sin\phi_l)^2 -p.
\ee
The squares can be linearized by a Hubbard-Stratonovich transformation.
This results in
\be
\langle |\tilde D_q |^{2p}\rangle
 &=&\int \frac{d\alpha d\beta}{2\pi} e^{-\alpha^2/2-\beta^2/2-\omega/2} 
 \int \prod_{k=1}^{2p} \frac{d\phi_k d\theta_k}{2\pi^2} e^{iq(\phi_1+\cdots +\phi_p)} e^{-iq(\theta_1+\cdots +\theta_p)} 
\nn \\ &&\times e^{
-\omega^{1/2}\alpha(\sum_{k=1}^p\cos \theta_k  +\sum_{k=1}^p \cos\phi_l)
-\omega^{1/2}\beta(\sum_{k=1}^p\sin \theta_k  +\sum_{k=1}^p \sin\phi_l)}.
\label{mom2pe2}
\ee
The sines and cosines can be added as
\be
\alpha\cos \phi +\beta\sin\phi =\sqrt{\alpha^2+\beta^2} \cos(\phi+\phi_0),
\ee
where
\be
\cos\phi_0 = \frac \alpha{\sqrt{\alpha^2+\beta^2}},
\ee
and the same for the $\theta$-variable. After shifting $\phi$ and 
$\theta$ by $-\phi_0$ we obtain
\be
\langle |D_q |^{2p}\rangle
&=&\int \frac{d\alpha d\beta}{2\pi} e^{-\alpha^2/2-\beta^2/2-\omega/2} \nn\\
&&\times
\left [ \int \frac{d\phi}{2\pi} e^{iq \phi -\omega^{1/2} \sqrt{\alpha^2+\beta^2}\cos \phi} \right ]^p 
\left [ \int \frac{d\theta}{2\pi} e^{-iq \theta -\omega^{1/2} \sqrt{\alpha^2+\beta^2}\cos \theta }\right ]^p 
\nn \\
&=& 
 e^{-\omega/2} \int_0^\infty r dr e^{-r^2/2} 
\left [ \int d\phi e^{iq \phi -\omega^{1/2} r\cos \phi} \right ]^{2p} 
\nn\\
&=& 
 e^{-\omega/2} \int_0^\infty r dr e^{-r^2/2} 
[I_q(\omega^{1/2} r)]^{2p}. 
\label{mom2pe4}
\ee
For $p=1$ the integral can be evaluated analytically
\be
 e^{-\omega/2} \int_0^\infty r dr e^{-r^2/2} 
[I_q(\omega^{1/2} r)]^{2} =e^{\omega/2}I_q(\omega). 
\label{mom2pe1}
\ee
\begin{figure}[t!]
\includegraphics[width=8.0cm]{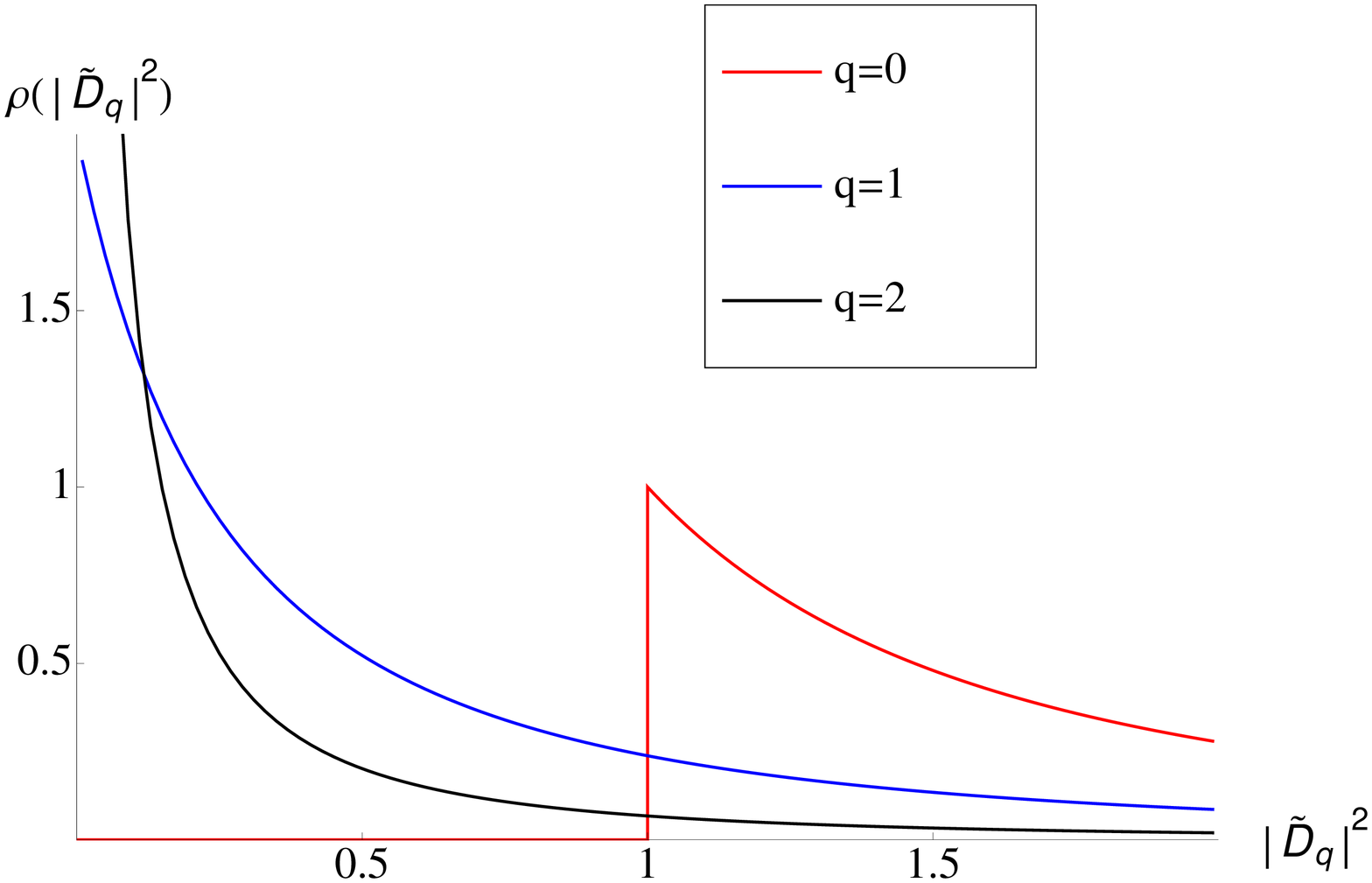}
\includegraphics[width=8.0cm]{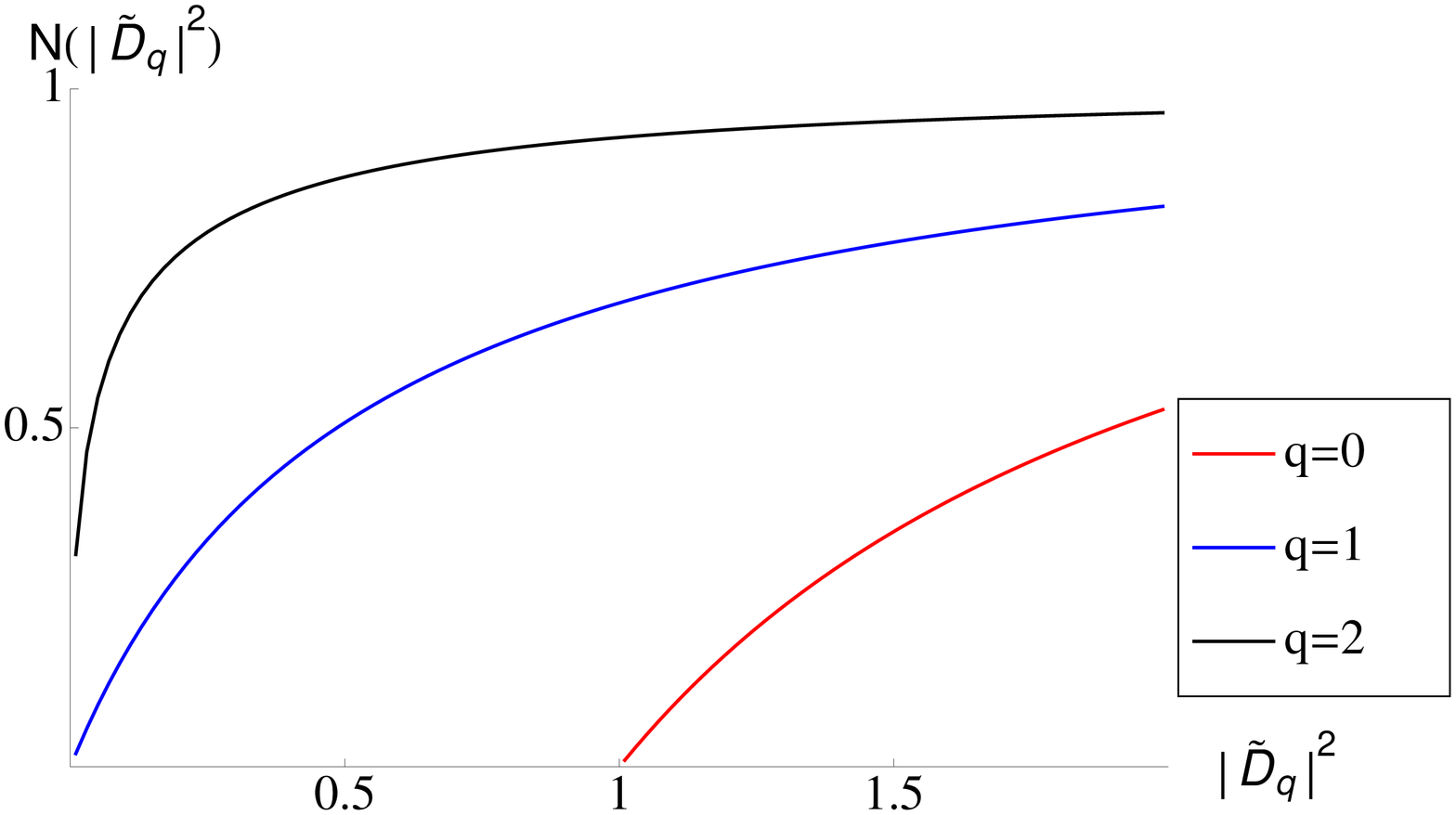}
\caption{The distribution (left) and the cumulative distribution (right)
of the absolute value squared for the canonical 
determinants $\tilde D_q$. Results are given for $q=0$, $q=1$ and $q=2$.}
\end{figure}

The normalized moments are given by
\be
\frac{\langle |\tilde D_q |^{2p}\rangle }{\langle |\tilde D_q |^{0}\rangle } 
&=&  
 \int_0^\infty r dr e^{-r^2/2} 
[I_q(\omega^{1/2} r)]^{2p} .
\label{mom2pe6}
\ee
The sum over $p$ in  Eq. (\ref{dens}) can be evaluated 
analytically after inserting the expression (\ref{mom2pe6}) 
for the moments. This results in the average density
\be
\rho_q(x)
&=& \frac 1{2\pi}\int_{-\infty}^\infty ds \int_0^\infty rdr e^{-r^2/2}
e^{-is(I_q(\omega^{1/2} r))^2 }
e^{ isx} \nn\\
&=&  \int_0^\infty rdr e^{-r^2/2} \delta (I_q(\omega^{1/2} r))^2-x) \nn\\
&=& \left . \frac{y e^{-y^2/2\omega}}
{2\omega I_q'(y) I_q(y) } \right |_{I_q(y)=\sqrt x}.
\ee
This can also be written as
\be
P\Big [[I_q^2]^{-1}[|\tilde D_q|^2] \in [x, x+dx] \Big ] = \frac x\omega e^{-x^2/2\omega}dx,
\ee
with $[I_q^2]^{-1}$ the inverse of the function $ x \to I_q^2(x)$. This is a
variant of a log-normal distribution which can be seen by
keeping only the exponential
factor in the asymptotic expansion of the Bessel functions. 
For  large
determinants we thus find 
\be
P\Big [\log|\tilde D_q| \in [x, x+dx] \Big ] \sim \frac x\omega e^{-x^2/2\omega}dx.
\ee
Another useful quantity is the cumulative distribution of the
canonical determinants. It is given by
\be
N_q(y) &=& \int_ 0^y \rho_q(y) dy \nn \\
     &=&\frac 1\omega \int_0^{\bar r(y)} rdr e^{-r^2/2\omega} \nn \\
  &=& 1- e^{-\bar r^2(y)/2\omega} 
\ee
with $ I_q(\bar r(y)) = \sqrt y$. For large $q$ we have that 
$I_q(x)\sim (x/2)^q/q!$ so that $\bar r \sim 2 qy^{1/2q}/e $ resulting in 
the cumulative distribution
\be
N(y) \sim 1- e^{-2q^2y^{1/q}/\omega e^2}.
\ee

\section{Conclusions}
\label{sec:conc}

For a fixed non-zero quark charge the canonical determinants in QCD take 
complex values. The average of these canonical determinants is the 
corresponding canonical partition function, which  
is a real and positive number. To address the cancellations which 
take place in forming the canonical partition functions  we have 
computed the distribution of the canonical determinants by means of  
chiral perturbation theory. In the limit of a dilute pion gas, the 
result simplifies to an expression in terms of modified 
Bessel functions. There are strong cancellations between the real and 
imaginary parts of the canonical
determinants which lead to an exponential suppression, 
$\exp(-q(m_N/3-m_\pi/2))$, with 
respect to the average magnitude. The magnitude is strongly fluctuating as well
with a distribution that 
in the low-temperature limit   
 is given by a variant of the log-normal distribution.
Moreover, we have evaluated the canonical partition
functions at non-zero isospin density.

Our results were obtained by means of chiral perturbation theory in the dilute limit, 
and as demonstrated the analytical form agrees qualitatively with lattice QCD results over 20 
orders of magnitude. The only caveat is that we used the arguments of the Bessel functions
as a fitting parameter and that the value of the pion mass is outside
the domain where chiral perturbation theory can be applied reliably. Consistent with the evaluation of the absolute value squared of the 
determinants in the dilute limit, the agreement is better for small $q$.
Further lattice studies with lighter quarks are needed for a full quantitative test of the 
analytic results. For such a test additional terms from the low temperature expansion 
should be included for the larger $q$ values.



We have also determined the contribution to the canonical determinants from the 
mean field saddle point corresponding to the Bose
condensed phase. We find an exponential suppression for small $q$ which turns into a 
Gaussian tail for large $q$. This effect becomes more relevant for larger values
of $q$.

It would be most interesting to determine the distribution of the canonical determinants 
in improved lattice simulations with light quarks. This would allow for a quantitative 
test of the analytical predictions and lead to a better understanding of the large 
$q$ behavior of the canonical determinants.  
\vspace{3mm}

\noindent
{\bf Acknowledgments:} 
This work was supported by U.S. DOE Grant No. DE-FG-88ER40388 (JV), the Austrian Science Fund FWF Grant Nr.~I 1452-N27 (CG), FWF DK W1203 ``Hadrons in Vacuum, Nuclei and Stars'' (H-PS), The U.S.~National Science Foundation CAREER grant PHY-1151648 and the {\sl Sapere Aude} program of The Danish Council for Independent Research (KS).


\renewcommand{\thesection}{Appendix \Alph{section}}
\setcounter{section}{0}
 
\section{Isospin chemical potential from the canonical partition functions}
\label{app:mufromZq}

The isospin chemical potential corresponding to the canonical partition
function (\ref{I-ratio}) is given by
\be
\mu_I(T,n_I) &=&\frac 12( -T \log Z_{q+2} +T \log Z_q)\nn\\ 
&=&-\frac T2 \log \frac{I_{(q+2)/2}(\omega)}{I_{q/2}(\omega)}.
\ee
In the thermodynamic limit the chemical potential at fixed isospin density 
is given by
\be
\mu_I(T, n_I)&=& -\frac T 2\lim_{q\to \infty } \log \frac{I_{(q+2)/2}(q \tilde \omega )}{I_{q/2}(q \tilde \omega)}
\ee
with
\be
\tilde \omega = \frac \omega q =\frac { m_\pi^2T }{n_I \pi^2}K_2(m_\pi/T),
\ee
and $n_I$ is the isospin charge density
\be
n_I = \frac {q}{V_3}.
\ee

 For large $q$ we can use the uniform approximation for modified Bessel functions
\be
I_q(q z) \sim \frac 1{\sqrt{2\pi q }}\frac {e^{q\eta}}{(1+z^2)^{1/4}}
\ee
with
\be 
\eta = \sqrt{1+z^2} +\log (z/2)  -\log(\frac 12+ \frac 12 \sqrt{1+z^2}).
\ee
In the low temperature limit, the argument of the modified Bessel functions is small so that
\be
\eta \sim 1 +\frac 14 z^2  +\log(z/2),
\ee
and
\be
I_\nu(\nu z) \sim \frac {(z/2)^\nu}{\sqrt{2\pi\nu}} e^{\nu(1+\frac 14 z^2) -\frac 14 z^2}.
\ee
For large $q$ and low temperatures we thus have
\be
I_{q/2}(q\omegat) &\sim& \frac {\omegat^{q/2}}{\sqrt{\pi q}} e^{q/2 +\omegat^2(q/2 -1)},\nn \\
I_{(q+2)/2}(q\omegat) &\sim&\frac {(\omegat q/(q+2))^{(q+2)/2}}{\sqrt{\pi (q+2)}} e^{(q+2)/2 +\omegat^2 q^3/2(q+1)^2 }.
\ee
In the large $q$ limit $I_{(q+2)/2}(q\omegat)$ simplifies to
\be
I_{(q+2)/2}(q\omegat) &\sim&\frac {\omegat^{q/2}}{\sqrt{\pi q}} 
e^{(q+2)/2 +\omegat^2q /2  }.
\ee
This results in the ratio
\be
\frac{I_{(q+2)/2}(q\omegat)}{I_{q/2}(q\omegat)} \sim \omegat e^{\omegat^2/2}.
\ee
For $\omegat \ll 1$ we obtain
\be
\frac{I_{(q+2)/2}(q\omegat)}{I_{q/2}(q\omegat)} \sim \omegat .
\label{uniform}
\ee
This results  in the chemical potential
\be
\mu_I(T,n_I) &\approx& -\frac T2 \log \omegat   \nn\\
     &=& \frac {m_\pi}2 -\frac T2  \log\left [ \frac{m_\pi^{3/2} T^{3/2}}{n_I \sqrt{2\pi^3}} \right ]  .
\label{mui-high}
\ee
In
 order to occupy the zero momentum states we need that $\mu > m_\pi/2$. The critical temperature for the formation
of a pion condensate is thus given by the relation $\mu_I (T_c, n_I)=m_\pi/2$. 
In the low temperature limit  the critical temperature in the $n=1$ approximation 
is thus given by
\be
T_c = 2^{1/3} \pi \frac {n_I^{2/3}}{m_\pi},
\label{crit}
\ee 
which up to the proportionality constant agrees with the result first obtained
by Einstein \cite{einstein} (See  \ref{app:B}).
It should be noted that  
for $(2mT)^{3/2}/n_I \sim O(1)$ the sum over $n$ in Eq. (\ref{zex}) cannot be truncated to
its first term.
A  calculation of the critical temperature that includes all terms is given in 
 \ref{app:B}.

\section{Critical Temperature for an Ideal Bose Gas}
\label{app:B}

In this Appendix we determine the critical temperature for a noninteracting Bose gas (see for example \cite{einstein,vosk})
and
explain that the low temperature approximation gives the correct scaling behavior but
does not reproduce the proportionality constant.

For density $n_I$,  the critical temperature is given by the condition
\be
\frac 12 n_I = \left .\int \frac{ d^3p}{(2\pi)^3} 
\frac 1{e^{\beta(\sqrt{p^2+m^2_\pi} -2\mu)}-1} \right |_{\mu =m_\pi/2}.
\ee
In the nonrelativistic approximation, this simplifies to
\be
\frac 12 n_I &=&  \frac 1{2\pi^2} \int p^2 dp 
 \frac 1{e^{\beta p^2/2m }-1} \nn\\ 
&=&  \frac 1{2\pi^2} (2mT)^{3/2}\int x^2 dx 
 \frac 1{e^{x^2}-1}  .
\ee
The integral can be evaluated as
\be
\int x^2 dx  \frac 1{e^{x^2}-1}
= \frac 14 \sqrt \pi \zeta(3/2) \approx 1.15758.
\ee
The numerical constant obtained this way differs from 
the constant in (\ref{crit}). However, if we make the
same approximation as in the derivation of Eq. (\ref{crit}), namely 
replacing the integral by
\be
\int_0^\infty dx \frac{x^2}{e^{x^2}-1} \to  \int_0^\infty dx x^2 e^{-x^2},
\ee
which corresponds to only keeping the $n=1$ term, we obtain
\be
\frac 12 n_I 
&=&  \frac 1{2\pi^2} (2mT)^{3/2}\int x^2 e^{-x^2} dx  \nn \\
&=& \left( \frac{ mT}{2\pi} \right )^{3/2}.
\ee
This results in the  expression (\ref{crit}) for the critical temperature.

\section{Canonical Partition Function at fixed isospin charge from mean field chiral perturbation theory}
\label{condensed}

When the isospin chemical potential is larger than half the pion mass, $\mu > m_\pi/2$, the grand 
canonical partition function enters in a phase in which the negatively charged
pions have condensed. In this case the mean field free energy depends on the
chemical potential and the mean field partition function is given by \cite{KSTVZ}
\be
Z(\mu) = e^{2V_4 F^2 \mu^2(1 +m_\pi^4/16\mu^4)}.
\ee
For $\mu \le m_\pi/2$ the mean field partition function is $ \mu $-independent, and for the entire
range of $\mu$ it can be written as
\be
Z(\mu) 
&=& e^{4V_4 F^2 (\mu^2 -m_\pi^2/4 )^2\theta(\mu^2-m_\pi^2/4)/2\mu^2+V_4 m_\pi^2 F^2}.
\ee
For $\mu$ close to $m_\pi/2$ it can be approximated by
\be
Z(\mu) 
&=& e^{8V_4 F^2 (|\mu| -m_\pi/2 )^2\theta(|\mu| -m_\pi/2) +V_4 m_\pi^2 F^2}.
\label{ztreshold}
\ee
For low temperatures, the corresponding canonical partition function
is given by
\be
\frac{Z_q}{Z_{q=0}} 
&=&  e^{-q^2/32 V_4F^2 T^2  -|q|m_\pi/2T}.
 \ee
This can be seen by evaluating
\be
\frac{Z(\mu)}{Z(\mu=0)} = \sum_q Z_q e^{\beta \mu q}.
\ee
For $|\mu| < m_\pi/2$ the sum over $q$ is dominated by the $q =0$ term so that
the partition function does not depend on $\mu$. For $\mu > m_\pi/2$, we
can do a saddle point approximation in $q$. This results  in the
partition function (\ref{ztreshold}). For $\mu < -m_\pi/2$, we find a saddle
point at negative $q$ again reproducing  (\ref{ztreshold}).
The free energy is an even function of $\mu$ so that the  canonical partition functions
are even in $q$ as well. In the mean field approximation the distinction between even and
odd $q$ has been lost and we do not find that $Z_q$ vanishes for odd $q$.

The canonical partition function can 
 be expressed in terms of the isospin density $n_I= q/V_3$, 
\be
\frac{Z_q}{Z_0} 
&=&  e^{-V_3 n_I^2/32 F^2 T    - |n_I| V_3 m_\pi/2T }.
\ee
This is the partition function of a repulsive Bose gas with vacuum energy density given by \cite{KSTVZ}
\be
E_0 = \frac{n_I^2}{32 F^2} + \frac  12 n_I m_\pi  .
\ee

In this case, the chemical potential for positive $q$ is given by
\be
\mu_I &=& -\frac T2 (\log Z_{q+2} - \log Z_q)\nn\\
 &=& \frac {m_\pi}2 +\frac q{16V_3F^2}.
\ee
Both $E_0$ and $\mu_I$ have been studied in lattice simulations
\cite{Detmold:2012wc} where qualitatively the same behavior was found.

If we create a density $n_0$ at zero temperature in the grand canonical ensemble,
so that
\be
n_0 = 16 F^2(\mu - m_\pi/2),
\ee
and then we heat the sample in the canonical ensemble at this density.
Then critical temperature from this mean field result is thus  given by (see (\ref{crit}))
\be
T_c &\sim& \frac {n_0^{2/3}}{m_\pi}\nn \\
&=& \frac {(\mu-m_\pi/2)^{2/3}}{m_\pi}.
\ee
 


\end{document}